\documentclass[pra,twocolumn,showpacs,amsmath,amssymb,superscriptaddress]{revtex4-1}
\usepackage{psfrag,graphicx}
\usepackage{amsfonts,amssymb,amsmath}      
\usepackage{graphicx}
\usepackage{dcolumn}
\usepackage{bm}
\usepackage{float}
\usepackage{hyperref}
\usepackage{color}
\usepackage{epstopdf}
\usepackage{mathtools}

\newcommand{\nn}{\nonumber\\}
\newcommand{\ex}[1]{\langle{#1}\rangle}

\usepackage{enumitem}
\newcommand{\beq}{  \begin{equation} }
\newcommand{\eeq}{\end{equation}}

\begin{document}

\title{Adaptive estimation of a time-varying phase with power-law spectrum via continuous squeezed states}

\author{Hossein T. Dinani}
\affiliation{Department of Physics and Astronomy, Macquarie University, Sydney, NSW 2109, Australia} 
\affiliation{Facultad de F\'isica, Pontificia Universidad Cat\'olica de Chile, Santiago 7820436, Chile\\}
\author{Dominic W. Berry}
\affiliation{Facultad de F\'isica, Pontificia Universidad Cat\'olica de Chile, Santiago 7820436, Chile\\}

\date{\today}

\begin{abstract}
When measuring a time-varying phase, the standard quantum limit and Heisenberg limit as usually defined, 
for a constant phase, do not apply. 
If the phase has Gaussian statistics and a power-law spectrum $1/|\omega|^p$ with $p>1$, then the generalized standard quantum limit and Heisenberg limit
have recently been found to have scalings of $1/{\cal N}^{(p-1)/p}$ and $1/{\cal N}^{2(p-1)/(p+1)}$, respectively, where ${\cal N}$ is the mean photon flux.
We show that this Heisenberg scaling can be achieved via adaptive measurements on squeezed states.
We predict the experimental parameters analytically, and test them with numerical simulations.
Previous work had considered the special case of $p=2$.
\end{abstract}

\maketitle

\section{Introduction}
Estimating a phase imposed on an optical beam is an important task in quantum metrology, particularly for accurate distance measurement
(for example gravitational wave detection).
Typically the performance of these measurements is quantified in terms of the photon number, because increasing the strength of the beam will improve the accuracy.
Standard techniques use coherent states, and have accuracy that is limited due to the statistics of coherent states.
Alternatively one may use squeezed states or more advanced states to improve the accuracy, which was originally proposed by Caves in 1981 \cite{Caves81}.
The ultimate limit to the accuracy using arbitrary states is often called the Heisenberg limit.

There are two scenarios for phase measurement that can be considered~\cite{WisMil10}.
One is an interferometer with a phase shift in one arm, where both modes are treated quantum mechanically, and the total photon number is considered as a resource.
The other is the phase shift on a single mode, which is estimated via quadrature measurements.
That is, the phase is measured relative to a strong local oscillator, which is treated classically, and only the photon number in the mode with the phase shift is considered as the resource.
In this work we consider the second scenario.

Phase measurements are most easily analyzed when the phase is constant.
In that case, the resource is just the average photon number $\bar n$.
The standard quantum limit (SQL) on the mean-square error (MSE) becomes $1/(4\bar n)$ \cite{Leon95}, and the Heisenberg limit becomes $1.89/\bar n^2$ \cite{Pegg90}.
(These are asymptotic scalings ignoring higher-order terms.)
There was much debate over the ultimate limits to phase measurement \cite{Bollinger96,Yurke86,Sanders95,Ou96,Zwierz10,Rivas12,Luis13,Luis13b,Anisimov10,Zhang13},
but the bounds have recently been proven \cite{Tsang12,Gio121,Berry12,Hall12,Nair12,Gio122,Hall12b}.
In the case of a constant phase, the analysis is simplified by the fact that there is an ideal canonical measurement which will yield the highest accuracy \cite{Leon95}.

In many applications, the quantity which one would wish to measure is varying in time, so the analysis for a constant phase no longer holds.
Some examples are:
\begin{enumerate}[noitemsep,topsep=0pt,parsep=0pt,partopsep=0pt,leftmargin=12pt]
\item For gravitational wave detection the signal of course varies in time.
\item Interferometers that are being developed for inertial sensors, with applications in seismology \cite{Collette15}.
\item The Gravity Recovery and Climate Experiment (GRACE) Follow-On mission is planned to include a laser interferometer for distance measurement \cite{Schutze16}.
\item Real-time phase measurement can be used to lock an interferometer that is being used for another purpose (such as photonic quantum logic).
\end{enumerate}

For some applications, such as gravitational wave detection, a particular type of signal is expected, so matched filtering can be used.
For more general measurement problems, the shape of the signal is not known in advance.
Instead, measurements may be performed with the only assumption on the signal being its spectrum.
A common assumption is that the spectrum scales as $1/|\omega|^p$ for $p>1$~\cite{DominicPRA02,DominicPRA06,DominicE06,DominicPRL13,DominicPRX}.
That is the case for binary inspiral gravitational waves \cite{Thorne}.
There are also many other situations that result in a power-law spectrum~\cite{Allan}.
For example, a random walk in frequency will result in a phase varying with $p=4$.

Because the photon number depends on the total time, and will go up indefinitely for a continuous measurement, it is better to quantify the resource by the average photon flux, ${\cal N}$.
To analyze this problem, it is necessary to choose a particular form of variation for the phase.
An early analysis considered phase that is varying as a Wiener process, and analyzed adaptive measurements using a squeezed state~\cite{DominicPRA02}.
In that work a broadband analysis of the squeezing was used without taking into account the photon flux resulting from the squeezing and found $1/{\cal N}^{2/3}$ scaling for the MSE. However, the photon flux for broadband squeezing is unbounded. 
A more advanced analysis in \cite{DominicPRA06} rectified this by treating the more difficult problem of narrowband squeezing, and found slightly poorer scaling of $1/{\cal N}^{5/8}$.
This analysis was further refined in \cite{DominicE06}, which found that the original scaling of $1/{\cal N}^{2/3}$ could in fact be obtained when the narrowband nature of the squeezing was  properly taken into account.

Up to this point these were just examples of measurements, and it was unknown what an equivalent of the Heisenberg limit for a varying phase would be.
This question was addressed in Ref.~\cite{DominicPRL13}, which showed that for \emph{squeezed states} no better scaling of the MSE could be obtained regardless of the measurement technique.
In fact, it showed 
for a general power-law spectrum with $p>1$,
the bound is scaling as $1/{\cal N}^{2(p-1)/(p+1)}$.
This scaling can therefore be regarded as a Heisenberg limit.
In the case of a Wiener process, $p=2$, this result yields the scaling found for adaptive measurements.
Reference \cite{DominicPRL13} also found scaling of $1/{\cal N}^{(p-1)/p}$ for coherent states, which can be regarded as a SQL for a varying phase.
These results were made more general in Ref.~\cite{DominicPRX}, which showed that not only for squeezed states, but all possible quantum states, the lower bound on the MSE is scaling as $1/{\cal N}^{2(p-1)/(p+1)}$.

Reference \cite{DominicPRX} also considered a simplified measurement scheme in order to show that the scaling can, in principle, be achieved.
The scheme involved sampling the phase at a sequence of times using pulses of squeezed light and interpolating the phase in between those samples.
Although it was possible to analytically prove results for that technique, it would not be practical, because it would require ideal phase measurements, or at least extremely fast adaptive measurements.
In addition, it can be expected to be suboptimal because it only samples the phase, rather than measuring it at all times.
In that work there was a significant difference in the constant for the scaling between the lower bound and the measurement technique. It is desirable to close this gap and find the best possible measurement.

In this work we theoretically consider adaptive measurements on a continuous-mode squeezed state (rather than a pulsed squeezed state), and show that the scaling of $1/{\cal N}^{2(p-1)/(p+1)}$ can still be obtained.
We analytically predict how the measurement performs, and verify the prediction via numerical simulations. We obtain an improvement in the scaling constant over that of the pulsed measurement scheme in \cite{DominicPRX} for values of $p$ up to about $1.5$.
In addition, we describe a technique to more accurately simulate the measurements by integrating the stochastic differential equations over short intervals.
Using this technique we recalculate the results of \cite{DominicE06}, and give more accurate corrected results.

We start by giving the details of the time variation of the phase. In Sec.~\ref{adsqmeas} we explain the adaptive measurement scheme. This is followed by the details of the feedback phase in Sec.~\ref{feedph}. We then analytically find the scaling of the experimental parameters in Sec.~\ref{anscal}. These scalings are confirmed through numerical simulations in Sec.~\ref{sec-numerics}.

\section{System phase time variation}\label{phvar}
We consider a time-varying system phase $\varphi{\left(t\right)}$ which has statistics that are Gaussian and stationary.
Therefore the mean value of the phase $\left\langle {\varphi{\left( t \right)}} \right\rangle$
is independent of time, and its autocorrelation function $ \Sigma{\left( {{t_1},{t_2}} \right)} = {\left\langle {\varphi {\left( {{t_1}} \right)} \varphi {\left( {{t_2}} \right)}} \right\rangle}$
is a function of only $t_1-t_2$.
In the following we will express $\Sigma$ as a function of only a single argument, which is the time difference.
Moreover, we assume the spectral density of the process, defined as the Fourier transform of the autocorrelation function, 
\begin{equation}
\tilde \Sigma {\left( \omega  \right)} = \int_{ - \infty }^\infty  {\Sigma {\left( t \right)}\,{e^{ - i\omega t}}\,dt}\, ,
\end{equation}
has power law scaling  for large $\omega$, i.e.~${{{\tilde{\Sigma }}}_{}}{\left( \omega  \right)} \sim  \kappa^{p-1}/\! {|\omega|}^p$. The multiplicative factor $\kappa$ is a constant with units of frequency, and is the inverse of the characteristic time of the spreading of the process. To ensure that the spectrum is limited at $\omega=0$, we consider the spectral density to be \cite{{DominicPRX},{vanTrees}}
\begin{equation}\label{spectrum-eq}
\tilde\Sigma(\omega)=\frac{\kappa^{p-1}}{|\omega|^p+\Gamma^p}\, .
\end{equation}
Here $\Gamma$ is a constant and is the characteristic time for the relaxation of the phase towards zero \cite{DominicPRX}.

For $p=2$, the phase varies as an Ornstein-Uhlenbeck process, and is easy to generate \cite{{Wheatley},{Yonezawa}}.
For general $p$, a time-varying phase can be generated via a Fourier transform \cite{{Percival},{Shinozuka}}.
Here we describe the technique we used.
Taking the Fourier transform of the phase $\varphi(t)$ and calculating the two-frequency expectation value gives
\begin{align}
&\langle \tilde\varphi(\omega_1) \tilde\varphi^* (\omega_2) \rangle \nonumber \\
&\quad = \int^{\infty}_{-\infty} \int^{\infty}_{-\infty}  dt_1 \, dt_2 \langle \varphi(t_1) \varphi(t_2) \rangle e^{-i (\omega_1 t_1 - \omega_2 t_2)} \nn
&\quad = \int^{\infty}_{-\infty}\int^{\infty}_{-\infty} dT \, d\Delta \, \Sigma(\Delta) e^{-i \left[\frac 12 (\omega_1 + \omega_2)\Delta + (\omega_1 - \omega_2)T \right]} \nn
&\quad = 2\pi\, \delta(\omega_1-\omega_2) \int^{\infty}_{-\infty} d\Delta \, \Sigma(\Delta) e^{-i \left[\frac 12 (\omega_1 + \omega_2)\Delta\right]} \nn
&\quad = 2\pi\, \delta(\omega_1-\omega_2) \tilde\Sigma (\omega_1)\, .
\end{align}
Here, we have used the change of variables $\Delta=t_1-t_2$, $T=(t_1+t_2)/2$, and in the last line we have replaced ${\left(\omega_1+\omega_2\right)}/2$ by $\omega_1$ because of the delta function $\delta{\left(\omega_1-\omega_2\right)}$.
Note also that, because the phase $\varphi(t)$ is real, $\tilde\varphi(-\omega)=\tilde\varphi^*(\omega)$. As a result, we can write the Fourier transform of the phase in the form
\begin{equation}
\tilde\varphi(\omega) = \sqrt{2\pi\tilde\Sigma(\omega)}\,\zeta(\omega)\, ,
\end{equation}
where $\zeta(\omega)$ has the correlations
\begin{equation}
\langle \zeta(\omega_1) \zeta^*(\omega_2) \rangle = \langle \zeta(\omega_1) \zeta(-\omega_2) \rangle = \delta(\omega_1-\omega_2)\, .
\end{equation}
Taking the inverse Fourier transform of $\tilde \varphi{\left({\omega}\right)}$ we obtain
\begin{equation}\label{xt}
\varphi {\left( t \right)}=\frac{1}{2\pi }\int_{{-\infty}}^{{\infty}}{d\omega\, \sqrt{2\pi {{{\tilde{\Sigma }}}_{}}{\left( \omega  \right)}}\zeta {\left( \omega  \right)}{{e}^{i\omega t}}}\, .
\end{equation}
Calculating the correlation function we obtain
\begin{align}\label{correlationf}
  \left\langle \varphi \left( t+\tau  \right)\varphi \left( t \right) \right\rangle 
&=\frac{1}{2\pi }\int_{-\infty}^{\infty}d\omega_1 \int_{-\infty}^{\infty}d{\omega }_2\sqrt{{{{\tilde{\Sigma }}}_{}}\left( \omega_1 \right){{{\tilde{\Sigma }}}_{}}\left( {{\omega_2 }} \right)}\nonumber \\ 
& \quad \times \left\langle \zeta \left( \omega_1  \right)\zeta \left( {{\omega _2}} \right) \right\rangle {{e}^{i\omega_1 (t+\tau )}}{{e}^{i{\omega_2 }t}} \nonumber \\
 &=\frac{1}{2\pi }\int_{-\infty}^{\infty}d\omega\, \tilde\Sigma(\omega)e^{i\omega \tau }\, .
\end{align}
This confirms that $\varphi(t)$ has power spectral density $\tilde\Sigma(\omega)$.

To generate this phase in our simulations we generate discretized complex white noise and use a discretized Fourier transform. We take $\zeta(\omega)$ to be approximated by 
\begin{equation}
\zeta {\left( \omega_k  \right)}\approx {\left( {{z}_{k,1}}+i{{z}_{k,2}} \right)}/\!\sqrt{2\delta \omega }\, ,
\end{equation}
where $z_{k,1}$ and $z_{k,2}$ are normally distributed random numbers with mean $0$ and variance $1$, that are independent except for $z_{k,1}=z_{-k,1}$, $z_{k,2}=-z_{-k,2}$. We approximate the integral in Eq.~\eqref{xt} by 
\begin{equation}
\varphi(t_n)\approx \frac{1}{\sqrt{4\pi}}\sum_k\sqrt{\delta \omega}\sqrt{\tilde\Sigma(\omega_k)}\left(z_{k,1}+iz_{k,2}\right)e^{i\omega _kt_n}\, .
\end{equation}
Taking $t_n=n\,\delta t$, $\omega_k=k\,\delta \omega$ and $\delta \omega \,\delta t=2\pi/\!N$ the above equation becomes
\begin{align}
 \varphi(t_n)&\approx \frac{1}{\sqrt{2N\delta t}}\sum_k\sqrt{\tilde\Sigma(\omega_k)}\left( z_{k,1}+iz_{k,2} \right)e^{i2\pi nk/N} \nonumber \\ 
& \approx \frac{1}{\sqrt{2N\delta t}}\left[ \sum_{k=0}^{N-1}{\sqrt{\tilde\Sigma(\omega_k)}\left( {z'_{k,1}}+i{z_{k,2}} \right){{e}^{i2\pi nk/N}} }\right. \nonumber \\
&\quad \left. +\sum\limits_{k=0}^{N-1}\sqrt{\tilde\Sigma(\omega_k)}\left( z'_{k,1}-iz_{k,2} \right)e^{-i2\pi nk/N} \right] \nonumber \\ 
 & =\sqrt{\frac{2}{N\delta t}}\left[ \operatorname{Re}\left( \sum\limits_{k=0}^{N-1}\sqrt{\tilde\Sigma(\omega_k)}\left( z'_{k,1} \right)e^{-i2\pi nk/N} \right) \right. \nonumber \\
&\quad \left. -\operatorname{Im}\left( \sum_{k=0}^{N-1}{\sqrt{\tilde\Sigma(\omega_k)}\left( z_{k,2} \right)e^{-i2\pi nk/N}} \right) \right] ,
\end{align}
where $z'_{k,1}=z_{k,1}$ for $k \ne 0$ and $z'_{0,1}=z_{0,1}/2$.
This phase can be efficiently calculated via a Fast Fourier Transform.

\begin{figure}[t!]
\centering
\includegraphics[scale=0.55]{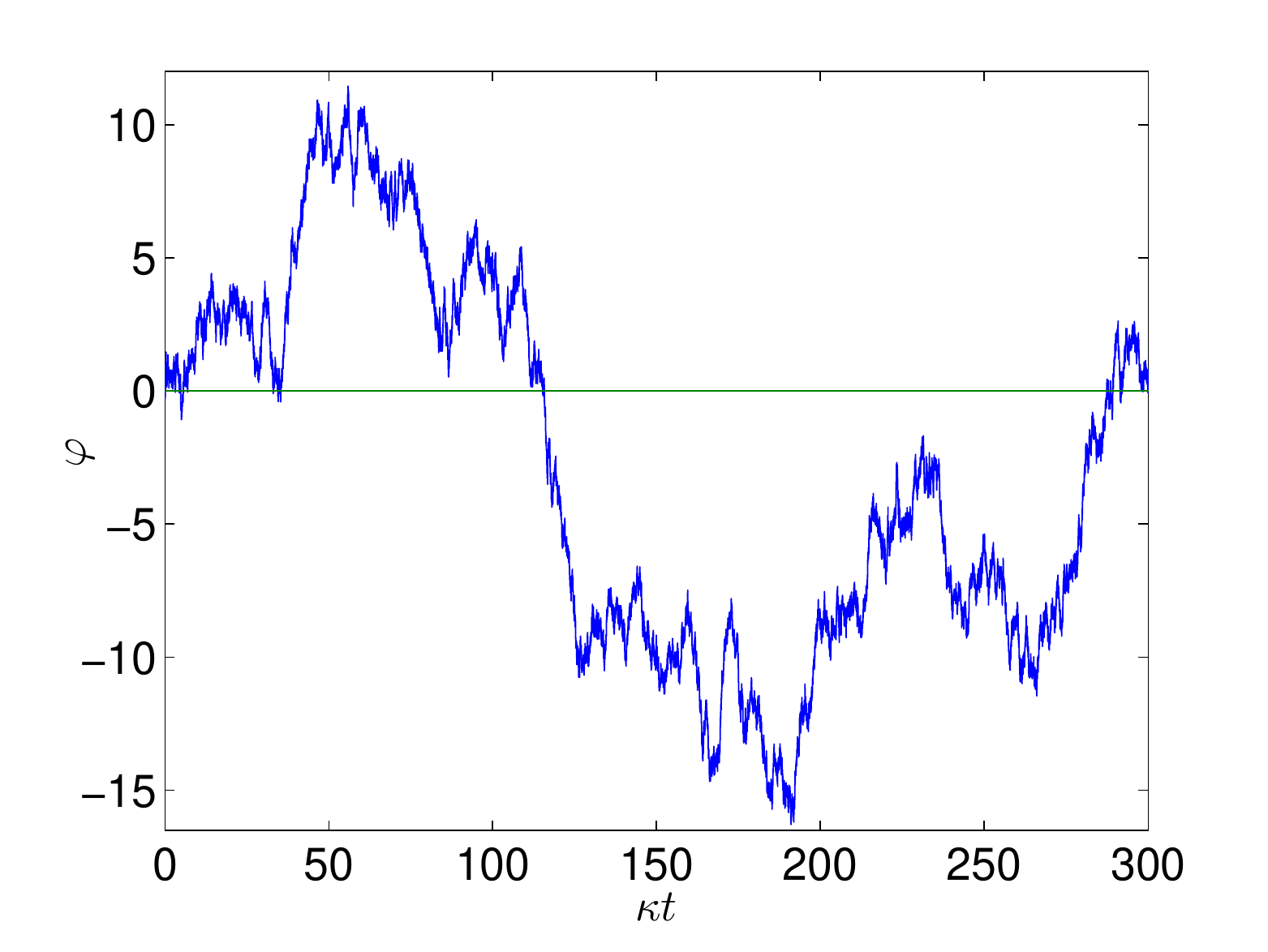}
\caption{A Gaussian random process with power law spectral density $\kappa^{p-1}/{\left({\omega^p+\Gamma^p}\right)}$, with $p=2$, $\Gamma/\kappa=10^{-3}$, and $\kappa\,\delta t=10^{-3}$. }
\label{fig-cnp2}
\end{figure}

Figure~\ref{fig-cnp2} shows the generated phase using the above equation for $p=2$ and $\Gamma/\kappa=10^{-3}$. As $p$ is increased the phase has less high-frequency variation.
In the next section, we give details of the measurement scheme for estimating such a varying phase.

\section{Adaptive measurement with squeezed states}\label{adsqmeas}
\begin{figure}[t]
\centering
\includegraphics[scale=1.95]{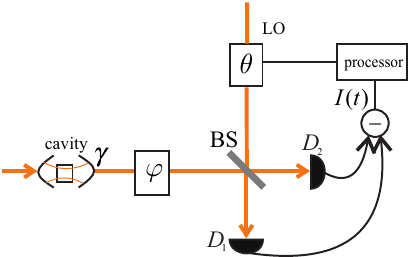}
\caption{The scheme for adaptive homodyne measurement of the phase $\varphi$ imposed on a squeezed coherent state generated by a cavity with decay constant $\gamma$. $D_1$ and $D_2$ are the photodetectors. $I(t)$ is the difference photocurrent between the two outputs of the 50/50 beam splitter (BS). The processor adjusts the phase of the local oscillator (LO) labeled by $\theta$ based on $I(t)$.}
\label{figcavity}
\end{figure}

We start by describing the form of the measurement, as depicted in Fig.~\ref{figcavity}, and provide a method to simulate the measurement that is improved over the one proposed in Ref.~\cite{DominicPRA06}.
The time-varying system phase $\varphi$ is probed by a continuous-mode squeezed coherent beam.
This beam is produced in an optical parametric oscillator \cite{Walls}, where a nonlinear medium inside a cavity is pumped with a coherent beam.
The cavity has a decay constant $\gamma$, and the light leaking out of the cavity provides the continuous beam.
Quadratures of the beam may be measured by combining it with a strong local oscillator (LO) on a 50/50 beam splitter.
The difference photocurrent in the outputs of the beam splitter then yields a measurement of the quadrature.
The LO also has a phase shift $\theta$ which may be controlled.
In a homodyne measurement the phase $\theta$ would be chosen to be close to $\varphi$.
In adaptive measurements, there is no prior knowledge of $\varphi$, but instead $\theta$ may be varied during the measurement based on the difference photocurrent \cite{Wiseman95,Armen} to approximate a homodyne measurement. \par
Let $\hat X$ and $\hat Y$ denote quadrature operators for the field just \emph{outside} the cavity but before the phase shift, and let $\hat x$ and $\hat y$ denote quadrature operators \emph{inside} the cavity.
The output photon flux from the cavity can be written in terms of the quadrature operators as \cite{DominicE06} 
\begin{equation}\label{phflux-eq}
4 \mathcal{N} = \ex{\hat X} ^2 + \ex{\hat Y} ^2 + \ex{:\Delta \hat X^2 + \Delta \hat Y^2:}\, .
\end{equation}
Here, $\ex{\hat X}=0$ and $\ex{\hat Y}=E$, where $E$ is the coherent amplitude of the field. The normally ordered variances of the quadratures are \cite{Collett,Gardiner} 
\begin{align}
&\ex{:\Delta \hat X^2:} = \ex{:\hat X^2:}  - \ex{:\hat X:}^2 = - \frac{\gamma \varepsilon}{1 + \varepsilon}\, , \nonumber \\
&\qquad \qquad \ex{:\Delta\hat Y^2:}  = \frac{\gamma \varepsilon}{1 - \varepsilon}\, ,
\end{align}
where $\varepsilon$ is a parameter related to the squeezing parameter $r$ according to 
\begin{equation}
\varepsilon  = \frac{e^r - 1}{e^r + 1}\, .
\end{equation}
This gives
\begin{equation}
\mathcal{N} = \frac{{{E^2}}}{4} + \frac{\gamma }{2}\sinh^2{\left( {\frac{r}{2}} \right)}\, .
\end{equation}

The Heisenberg equation of motion for the quadrature operators \emph{inside} the cavity can be written as \cite{Fizek,DominicPRA06}
\begin{align}
 \frac{{d \hat x}}{dt} &=  - \hat x \gamma (1 + \varepsilon )/2 + \sqrt \gamma  \hat \xi\, ,  \label{eq-xop}\\ 
 \frac{{d \hat y}}{dt} &=  - \hat y \gamma \left( {1 - \varepsilon } \right)/2 + \sqrt \gamma  \hat\eta\, . \label{eq-yop}
\end{align}
Here, $\hat \xi$ and $\hat \eta$ are the quadrature noise operators, and we have considered the squeezed quadrature to be $\hat x$. 
The phase $\varphi$ is imposed on the squeezed state before it combines on a 50/50 beam splitter with a LO which has phase $\theta$. The output quadrature at angle $\theta-\varphi$  is obtained as \cite{{DominicPRA06},{Fizek}}
\begin{equation}\label{eq-I}
\hat I = \cos \left( {\theta  - \varphi } \right) \left( {\sqrt \gamma \hat x - \hat \xi } \right) + \sin \left( {\theta  - \varphi } \right)\left( {\sqrt \gamma  \hat y + E - \hat \eta } \right)\, .
\end{equation}
This corresponds to the measured difference photocurrent in the output modes.

Because Eqs.~\eqref{eq-xop}, \eqref{eq-yop}, and \eqref{eq-I} 
are linear in the cavity quadratures $\hat x$ and $\hat y$, and the bath quadrature inputs 
$\hat \xi$ and $\hat \eta$,
they can be simulated exactly using classical variables~\cite{DominicPRA06}. That is, we can replace these operators, 
and $\hat I$, by real-valued variables with the same statistics, as determined by the Wigner 
function for the quantum fields~\cite{Gardiner}. We can write
\begin{align}
\frac{{dx}}{{dt}} &=  - x\gamma (1 + \varepsilon )/2 + \sqrt \gamma   \xi \, ,  \label{eq-x}\\ 
\frac{{dy}}{{dt}} &=  - y\gamma \left( {1 - \varepsilon } \right)/2 + \sqrt \gamma  \eta \, , \label{eq-y}\\
I &= \cos {\left( {\theta  - \varphi } \right)} {\left( {\sqrt \gamma  x - \xi } \right)} + \sin {\left( {\theta  - \varphi } \right)}{\left( {\sqrt \gamma  y + E - \eta } \right)} \, .\nonumber\\
\end{align}
Here, $\xi$ and $ \eta$ are Gaussian increments satisfying $\left\langle \xi(t)\xi (t') \right\rangle = \left\langle \eta(t)\eta(t') \right\rangle  = \delta(t-t')$.
One way to numerically integrate these equations is to directly discretize the equations over time steps of length $\Delta t$ \cite{DominicPRA06}. The method we describe here is to instead integrate the differential equations over a time step of length $\Delta t$. This method is still not exact because we assume that the system and controlled phases are constant over these time intervals. That is, the remaining approximation in the discretization is now in taking the phases to be constant over the time intervals.  Provided the time intervals are short, the approximation will be accurate, and it will be more accurate than the approximation without the integrals.

Integrating Eqs.~\eqref{eq-x} and \eqref{eq-y} we obtain
\begin{align}
x(t) &= {e^{\gamma \left( {1 + \varepsilon } \right)\left( {{t_0} - t} \right)/2}}{x_0} \nonumber \\
&\quad+ \sqrt \gamma  \int_{{t_0}}^t {du\, {e^{\gamma (1+\varepsilon)(u-t)/2}}\xi(u)}\, , \\
y(t) &= {e^{\gamma \left( {1 - \varepsilon } \right)\left( {{t_0} - t} \right)/2}}{y_0} \nonumber \\
&\quad+ \sqrt \gamma  \int_{{t_0}}^t {du\, {e^{\gamma (1-\varepsilon)(u-t)/2}}\eta(u)}\, ,
\end{align}
where $x_0$ and $y_0$ are the values of $x$ and $y$ at $t=t_0$.
To obtain the effect of a step from time $t_0$ to $t_1=t_0+\Delta t$ we integrate $I$ over this interval. Therefore, we need to integrate $\sqrt{\gamma}x-\xi$ and $\sqrt{\gamma}y-\eta$. 
We obtain 
\begin{align}
&\int_{t_0}^{t_1} dt\, \left[ \sqrt \gamma  x - \xi (t) \right] = {x_0}\frac{e^{ - r} + 1}{\sqrt \gamma }\left( {1 - {e^{ - \gamma \left( {1 + \varepsilon } \right)\Delta t/2}}} \right)\nn
&- (e^{-r}+1)\! \int_{t_0}^{t_1} \! du\,\xi(u)e^{\gamma (1+\varepsilon)(u-t_1)/2} + \int_{t_0}^{t_1} \! du  \,\xi(u)e^{-r} \, .
\end{align}
Similarly for $\sqrt{\gamma}y-\eta$ we obtain
\begin{align}
&\int_{t_0}^{t_1} dt\, \left[ \sqrt \gamma  y - \eta(t) \right]={y_0}\frac{{{e^r} + 1}}{{\sqrt \gamma  }}\left( {1 - {e^{ - \gamma \left( {1 - \varepsilon } \right)\Delta t/2}}} \right)  \nn
& - (e^r+1)\int_{t_0}^{t_1}  du\, \eta(u) e^{\gamma (1-\varepsilon)(u-t_1)/2} + \int_{t_0}^{t_1} du\, \eta (u)e^r \, .
\end{align}
We define
\begin{align}
 {\chi _x} &\coloneqq \int_{t_0}^{t_1} {du\,\xi{\left( u \right)}{e^{\gamma{\left( {1 + \varepsilon } \right)}\left( {u - {t_1}} \right)/2}}}\, , \\
{\chi _y} &\coloneqq \int_{t_0}^{t_1} {du\,\eta {\left( u \right)}{e^{\gamma \left( {1 - \varepsilon } \right)\left( {u - {t_1}} \right)/2}}}\, , 
\end{align}
and
\begin{equation}
{\psi _x} \coloneqq \int_{t_0}^{t_1} {du\,\xi{\left( u \right)}e^{-r}} ,\qquad {\psi _y} \coloneqq \int_{t_0}^{t_1} {du\, \eta(u)e^r}\, .
\end{equation}
In terms of these new variables the integral of $I$ can be written as 
\begin{align}
&\int_{t_0}^{t_1} {dt\,I} = \cos(\theta-\varphi)\left[ x_0\frac{ e^{-r} + 1}{\sqrt \gamma}\left( 1 - e^{-\gamma (1+\varepsilon)\Delta t/2} \right)\right.  \nonumber \\
&\quad \left. \vphantom{\frac{ e^{-r} + 1}{\sqrt \gamma}} -\left( e^{-r}+1 \right)\chi _x + \psi _x \right] + \sin\left( \theta - \varphi  \right)\left[ y_0\frac{e^r + 1}{\sqrt \gamma}\right.
\nonumber \\ & \quad \left. \vphantom{\frac{ e^r + 1}{\sqrt \gamma}} \times\left( 1 - e^{ - \gamma (1-\varepsilon)\Delta t/2} \right) - \left( e^r + 1 \right)\chi _y + \psi _y+ E\Delta t  \right]\, .
\end{align}
The expectation values of $\chi_\ell$ and $\psi_\ell$ for both $\ell=x, y$ are zero because $\xi$ and $\eta$ both have mean zero. Therefore the variances are
\begin{align}
\left\langle {\chi _x^2} \right\rangle  &= \int_{t_0}^{t_1} {du\,{e^{\gamma \left( {1 + \varepsilon } \right)\left( {u - {t_1}} \right)}}}  \nonumber \\
&= {\left( {{e^{ - r}} + 1} \right)}{\left( {1 - {e^{ - \gamma {\left( {1 + \varepsilon } \right)}\Delta t}}} \right)}/2\gamma\, , \\
\left\langle {\chi _y^2} \right\rangle  &= \int_{t_0}^{t_1} {du\,{e^{\gamma \left( {1 - \varepsilon } \right)\left( {u - {t_1}} \right)}}}  \nonumber \\
&= {\left( {{e^r} + 1} \right)}{\left( {1 - {e^{ - \gamma {\left( {1 - \varepsilon } \right)}\Delta t}}} \right)}/2\gamma\, , \\
 \left\langle {\psi _x^2} \right\rangle  &= {e^{ - 2r}}\Delta t\, ,\qquad \left\langle {\psi _y^2} \right\rangle  = {e^{2r}}\Delta t\, ,
\end{align}
and the covariances are
\begin{align}
\left\langle {{\chi _x}{\psi _x}} \right\rangle  &= \int_{t_0}^{t_1} {du\,{e^{\gamma \left( {1 + \varepsilon } \right)\left( {u - {t_1}} \right)/2}}{e^{ - r}}} \nonumber \\
&= {e^{ - r}}{\left( {{e^{ - r}} + 1} \right)}{\left( {1 - {e^{ - \gamma {\left( {1 + \varepsilon } \right)}\Delta t/2}}} \right)}/\gamma\, , \\
\left\langle {{\chi _y}{\psi _y}} \right\rangle  &= \int_{t_0}^{t_1} {du\,{e^{\gamma {\left( {1 + \varepsilon } \right)}{\left( {u - {t_1}} \right)}/2}}{e^r}}  \nonumber \\
&= {e^r}{\left( {{e^r} + 1} \right)}{\left( {1 - {e^{ - \gamma {\left( {1 - \varepsilon } \right)}\Delta t/2}}} \right)}/\gamma\, .
\end{align}
We also define
\begin{align}
m_{x}^{(1)} &\coloneqq \left( {{e}^{-r}}+1 \right)\left( 1-{e^{-\gamma \left( 1+\varepsilon  \right)\Delta t/2}} \right)/\sqrt{\gamma }\, , \\
m_{y}^{(1)} &\coloneqq \left( {{e}^{r}}+1 \right)\left( 1-{e^{-\gamma \left( 1-\varepsilon  \right)\Delta t/2}} \right)/\sqrt{\gamma }\, ,
\end{align}
and ${{\Omega }_{x}}={{\psi }_{x}}-{{\lambda }_{x}}{{\chi }_{x}}$, and ${{\Omega }_{y}}={{\psi }_{y}}-{{\lambda }_{y}}{{\chi }_{y}}$ in such a way that the covariances $\left\langle {{\Omega }_{x}}{{\chi }_{x}} \right\rangle $, and $\left\langle {{\Omega }_{y}}{{\chi }_{y}} \right\rangle $ are zero.
The appropriate values of $\lambda_x$ and $\lambda_y$ are
\begin{equation}
 {{\lambda }_{x}}=\frac{\left\langle {{\chi }_{x}}{{\psi }_{x}} \right\rangle }{\left\langle \chi _{x}^{2} \right\rangle }\, ,\qquad {{\lambda }_{y}}=\frac{\left\langle {{\chi }_{y}}{{\psi }_{y}} \right\rangle }{\left\langle \chi _{y}^{2} \right\rangle }\, .
\end{equation}
In terms of these scaling factors we can write
\begin{align}
x(t_1) &= e^{-\gamma \left( 1+\varepsilon  \right)\Delta t/2}x_0+\sqrt{\gamma}\chi_x\, , \label{x-eq}\\
y(t_1) &= e^{-\gamma \left( 1-\varepsilon  \right)\Delta t/2}y_0+\sqrt{\gamma}\chi_y\, , \label{y-eq} \\ 
I(t_1) &= I(t_0)+\cos\left( \theta -\varphi  \right)\left( m_x^{(1)}x_0+\Omega_x+m_x^{(2)}\chi_x \right) \nonumber \\
& \quad+\sin\left( \theta -\varphi  \right)\left( m_{y}^{(1)}y_0+E\Delta t+\Omega_y+m_y^{(2)}\chi_y \right) , \label{I-eq}
\end{align}
where $m_x^{(2)}=\lambda_x-e^{-r}-1$, and $m_y^{(2)}=\lambda_y-e^r-1$. \par
In our numerical simulations we used Eqs.~\eqref{x-eq}, \eqref{y-eq} and \eqref{I-eq} to improve the accuracy.
Next we explain how the controlled phase $\theta$ is determined from the difference photocurrent.

\section{Feedback phase}
\label{feedph}
To estimate the time-varying phase $\varphi$ we change the LO phase $\theta$ based on the difference photocurrent given in Eq.~\eqref{I-eq} during the course of the measurement. 
The LO phase could be updated by Bayesian updating \cite{{Pope},{Olivares},{Berni}} or based on the functions of the photocurrent record $A$ and $B$ described in Refs.~\cite{DominicPRA02,DominicPRA06}.
The Bayesian updating is highly numerically intensive for this problem. Moreover, it is shown in Ref.~\cite{DominicPRA06} that Bayesian updating gives only a few percent enhancement over the other method.
Therefore, we follow the method of Refs.~\cite{{DominicPRA02},{DominicPRA06}}.

The relevant information from the measurement record can be formulated in the following quantities \cite{{DominicPRA02},{DominicPRA06}}
\begin{align}
A(t) &= \int_{ - \infty }^t {{e^{\chi {\left( {u - t} \right)}}}{e^{i\theta }}I(u)\,du}\, , \label{eq-A}\\
B(t) &=  - \int_{ - \infty }^t {{e^{\chi {\left( {u - t} \right)}}}{e^{2i\theta }}\,du}\, ,  \label{eq-B}
\end{align}
where $1/\chi$ is a time constant for the weight $e^ {\chi {\left( u-t \right)}}$ given to the difference photocurrent at time $u$, $I(u)$.
The phase estimate at time $t$, $\breve \varphi (t) $, is obtained from $A(t)$ and $B(t)$ via
\begin{equation}\label{eq-phiest}
\breve{\varphi}(t) =\arg{\left( C (t) \right)}\, , \qquad C(t)=A(t)+\chi B(t) A^{*}(t)\, .
\end{equation}
However, it is found that using this phase estimate as the LO phase gives poor results \cite{DominicPRA02,DominicPRA06}.
This is because for very good estimates of the phase in the feedback, the results do not distinguish easily between the system phase and system phase plus $\pi$.
Therefore, many of the results are out by $\pi$ which results in a large MSE.
Thus, following the technique of previous works \cite{DominicPRA02,DominicPRA06} we set the LO phase to 
\begin{equation}\label{eq-thetadelta}
\theta(t) =\arg {\left( C^{1-\delta }(t)A^{\delta }(t) \right)}\, ,
\end{equation}
and find the optimal value of $\delta$ numerically. 

Note that changing the controlled phase by $\pi$ does not make any difference to the final MSE.
If we were only attempting to measure the phase at a final time, then it would not matter if there were errors of $\pi$ in the phase estimate at intermediate times, because errors of $\pi$ in the controlled phase do not adversely affect the results.
This means that it would be reasonable to use $\arg{\left(C(t)\right)}$ as the phase estimate at intermediate times, and use larger values of $\delta$ close to the final time.
That is the approach used in Ref.~\cite{Berry00}.
However, because we require accurate estimates of the phase at all times, we must be able to resolve the $\pi$ ambiguity at all times, and it is better to use the LO phase given in Eq.~\eqref{eq-thetadelta}. 

The parameters that we can vary to obtain the minimum MSE are the squeezing parameter $r$, the cavity decay $\gamma$, the time scale over which previous measurement results are used $1/\chi$, and the parameter $\delta$.
If we scale the time by $\kappa$, we obtain the dimensionless parameters $\mathcal{{N}}/\kappa$, $\gamma/\kappa$, and $\chi/\kappa$.
The parameters $r$ and $\delta$ are already dimensionless.
We consider arbitrary squeezing; in other words, we do not consider any limitations for the squeezing parameter $r$.
This is because we want to find the ultimate scaling obtained with this scheme regardless of the current technological status of sources of squeezing.
Except for the parameter $\delta$, it is possible to analytically find the scaling of the parameters with ${\cal N}/\kappa$ as we derive in the next section.

\section{Analytical scaling}
\label{anscal}
First we note that for the given photon flux $\mathcal{N}$, the average number of photons in the time scale $1/\chi$ is $\mathcal{N}/\chi$. 
For the coherent state, the MSE scales with the inverse of the average number of photons, i.e.~$\chi/\mathcal{N}$ \cite{DominicPRA06}.
In adaptive measurements the MSE obtained with a squeezed state is reduced by a factor of $e^{-2r}$, so scales as $\chi e^{-2r}/\mathcal{N}$ \cite{DominicPRA06}.

To find the scalings of the parameters and MSE in terms of $\mathcal{N}/\kappa$ the following conditions should hold:\vspace{4pt}\\
\textbf{Condition 1:} \textit{The mean-square variation of the system phase in the time scale over which previous measurement results is used, $1/\chi$, should be on the order of the MSE obtained with squeezing.}\vspace{4pt}\\
\textbf{Condition 2:} \textit{The squeezing parameter should be chosen in such a way that the noise from the antisqueezing component is not larger than the noise from the squeezed quadrature.}\vspace{4pt}\\
\textbf{Condition 3:} \textit{To observe the effect of squeezing the time scale over which previous measurement results is used, $1/\chi$, needs to be on the order of or longer than $e^r/\gamma$ \cite{DominicPRA06}.}\vspace{4pt}\\
\textbf{Condition 4:} \textit{The parameters $\gamma$ and $r$ should not correspond to a photon flux due to squeezing larger than ${\cal N}$.}\vspace{4pt}

For spectral density $\sim\kappa^{p-1}/|\omega|^p$ the mean-square variation in the system phase after time $t$, ${\langle {\left( \varphi (t) - \varphi (0)\right)}^2 \rangle}$, is $\sim(\kappa t)^{p-1}$ \cite{DominicPRX}. For Condition 1 to hold we therefore need
\begin{equation}\label{chi-eq}
{\chi }{e^{ - 2r}}/{{\mathcal{N}}}\sim{\left( {{\kappa }/{\chi }} \right)^{p - 1}}\, .
\end{equation}
This gives
\begin{equation}\label{chisc}
\chi \sim \kappa^{1-1/p} {\left({\mathcal{{N}}e^{2r}}\right)}^{1/p}\, ,
\end{equation}
and therefore for the MSE with squeezing we 
obtain
\begin{equation}
\label{mse1}
{\rm MSE}\sim{\left({\kappa e^{-2r}/{\cal N}}\right)}^{1-1/p}\, .
\end{equation}

The difference photocurrent in the output for the adaptive homodyne measurement can be written as~\cite{DominicPRA02}
\begin{align}
I(t)dt &= 2E\sin \left( {\varphi (t) - \breve\varphi (t)} \right)dt + \sqrt {{R_{\rm sq}}(t)} dW(t)\, ,\\
{R_{\rm sq}}(t) &= {\sin^2}{\left( {\varphi (t) - \breve\varphi (t)} \right)}{e^{2r}} + {\cos^2}{\left( {\varphi (t) - \breve\varphi (t)} \right)}{e^{ - 2r}}
\label{eq48},
\end{align}
where $dW(t)$ is an infinitesimal Wiener increment which satisfies $\langle dW(t) dW(t')\rangle = \delta(t-t') (dt)^2$.
The amplitude of the Wiener noise, $R_{\rm sq}$, consists of the squeezing and antisqueezing components.

If the estimated phase $\breve \varphi$ is close to the system phase $\varphi$, we can approximate $R_{\rm sq}$ by $ e^{2r}{\rm MSE}+e^{-2r}$.
If we were to increase $r$ without limit for any nonzero value of MSE, then the first term for antisqueezing would eventually dominate.
Condition 2 above means that $r$ is sufficiently small and the phase estimate is sufficiently accurate that the first term in Eq.~\eqref{eq48} is not dominating.
When $e^{-2r} \sim e^{2r}{\rm MSE}$ the antisqueezed component starts to give significant noise, and increased squeezing will only increase the error.
To not have the squeezing beyond this point, the strongest squeezing we can have is such that $e^{-4r}\sim {\rm MSE}$.
Using Eq.~\eqref{mse1}, we obtain
\begin{equation}
e^{-4r} \sim {\left({\kappa e^{-2r}/{\cal N}}\right)}^{1-1/p}\, .
\end{equation}
Solving for $e^r$ gives
\begin{equation}\label{er-eq}
e^r \sim {\left({\mathcal{N}/\kappa}\right)}^{(p-1)/(2p+2)}\, . 
\end{equation}
That enables us to obtain the scaling for the MSE as
\begin{equation}\label{MSE-eq}
{\rm MSE} \sim {\left({\kappa/\mathcal{N}}\right)}^{2(p-1)/(p+1)}\, ,
\end{equation}
which is the Heisenberg scaling from Ref.~\cite{DominicPRX}.
Similarly, using Eq.~\eqref{chisc} we obtain the scaling for $\chi/\kappa$ as
\begin{equation}\label{chi-eq2}
\chi/\kappa \sim {\left({\mathcal{N}/\kappa}\right)}^{2/(p+1)}\, .
\end{equation}

This equation also shows the relation between the time scale at which the local oscillator should be updated to the time scale of the phase variation. The local oscillator phase should be updated in such a way that its variation is not much more than the MSE. According to Condition 1, the system phase varies over time $1/\chi$ by an amount comparable to the MSE. This means that the local oscillator should be updated in shorter time intervals than $1/\chi$ in order to keep the local oscillator phase sufficiently close to the system phase.
From Eq.~\eqref{chi-eq2} we see that as ${\cal N}/\kappa$ (the number of photons in the time scale of the system phase variation) increases, the local oscillator should be updated more rapidly as compared to the variation of the system phase.

\begin{figure*}[tbh]
\centering
\includegraphics[scale=0.825]{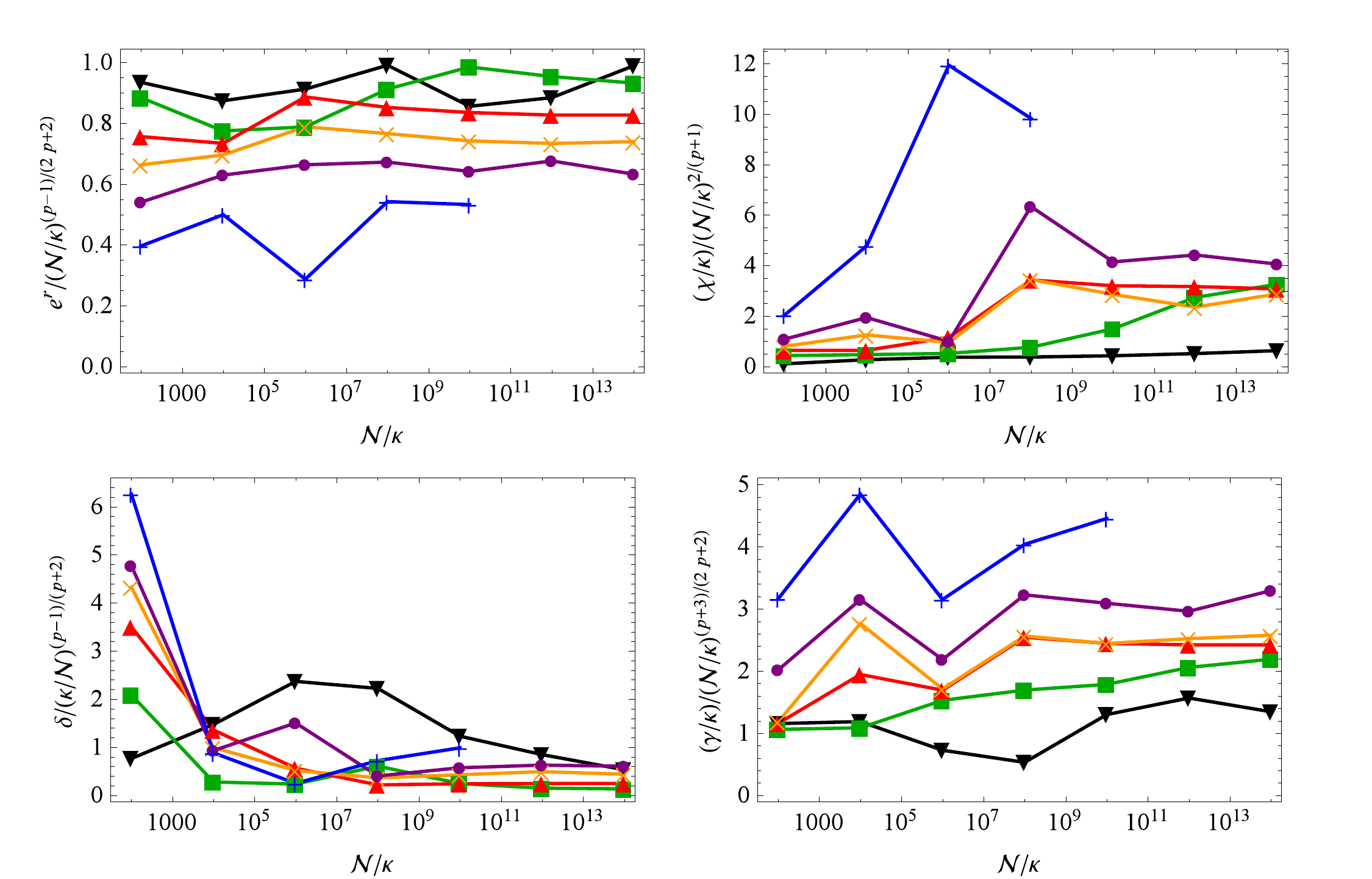}
\caption{The optimal values of the parameters $e^r$, $\chi$, $\delta$, and $\gamma$ vs.~$\mathcal{N}/\kappa$ for a range of values of $p$. Black line with $\blacktriangledown$: $p=1.25$, green line with $\blacksquare$: $p=1.5$, red line with $\blacktriangle$: $p=2$, yellow line with $\times$: $p=2.5$, purple line with \textbullet: $p=3$, and blue line with $+$: $p=4$.}\label{fig-4scalings}
\end{figure*}

So far, it is not guaranteed that this scaling can be reached, because there are also Conditions 3 and 4.
Condition 3, which is justified in Ref.~\cite{DominicPRA06}, gives us the inequality
\begin{equation}\label{gamineq-eq}
{\chi }{e^r}/{\kappa } < \gamma/\kappa \, .
\end{equation}
The smallest $\gamma$ this inequality can be satisfied with is
\begin{equation}\label{gamsca}
{\gamma }/{\kappa } \sim {\left( {\mathcal{N}}/\kappa \right)^{(p + 3)/(2p + 2)}}\, .
\end{equation}
This scaling of $\gamma$ will be acceptable provided it is not so large that it violates Condition 4.
Condition 4 implies that, using the equation for the photon flux Eq.~\eqref{phflux-eq}, we should have $\mathcal{N}>\gamma e^r$.
It turns out that, using the scalings in Eq.~\eqref{gamsca} and \eqref{er-eq}, we obtain $\gamma e^r\sim \mathcal{N}$, which does not violate Condition 4.

Note that for $p=2$ the scalings found here reproduce the scalings found in Ref.~\cite{DominicE06}.
We have not found an analytical way to determine the scaling of the parameter $\delta$.
In the next section we numerically find the scaling of $\delta$ and confirm the scalings of the other parameters we found in this section.

\begin{figure}[tbh]
\centering 
\includegraphics[scale=0.7]{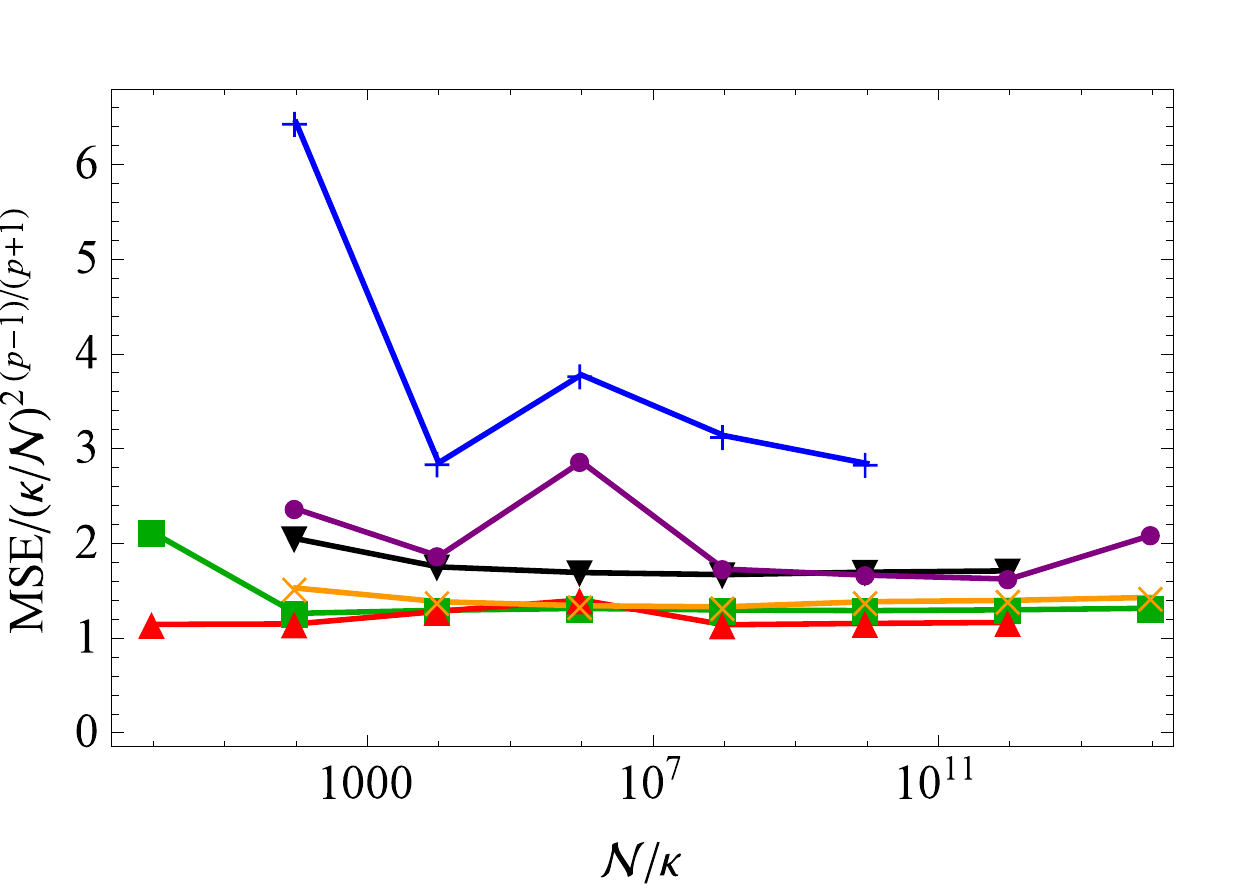}
\caption{The scaled MSE vs.~$\mathcal{N}/\kappa$ for a range of values of $p$. Black line with $\blacktriangledown$: $p=1.25$, green line with $\blacksquare$: $p=1.5$, red line with $\blacktriangle$: $p=2$, yellow line with $\times$: $p=2.5$, purple line with \textbullet: $p=3$, and blue line with $+$: $p=4$.}\label{fig-var}
\end{figure}

\begin{figure}[tbh]
\centering
\begin{picture}(600,325)
\put(30,180){\includegraphics[scale=0.6]{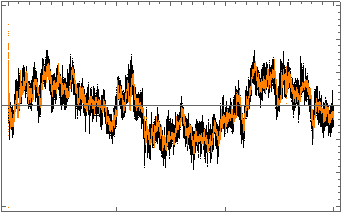}}
\put(30,20){\includegraphics[scale=0.6]{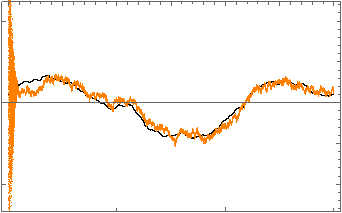}}
\put(130,159){$\chi t$}
\put(10,242){\rotatebox{90}{$\varphi$}}
\put(118,293){$p=1.5$}
\put(9,181){$-0.3$}
\put(23,242){$0$}
\put(16,302){$0.3$}
\put(93,171){$100$}
\put(159,171){$200$}
\put(224,171){$300$}
\put(130,0){$\chi t$}
\put(2,83){\rotatebox{90}{$\varphi$}}
\put(118,133){$p=3$}
\put(0,33){$-0.002$}
\put(23,83){$0$}
\put(8,132){$0.002$}
\put(93,11){$100$}
\put(159,11){$200$}
\put(224,11){$300$}
\put(40,295){(a)}
\put(40,135){(b)}
\end{picture}
\caption{System phase (black line) and the estimated phase (orange line) for $\mathcal{N}/\kappa=10^8$. The values of $p$ are (a) $p=1.5$ and (b) $p=3$.}\label{fig-cnest}
\end{figure}
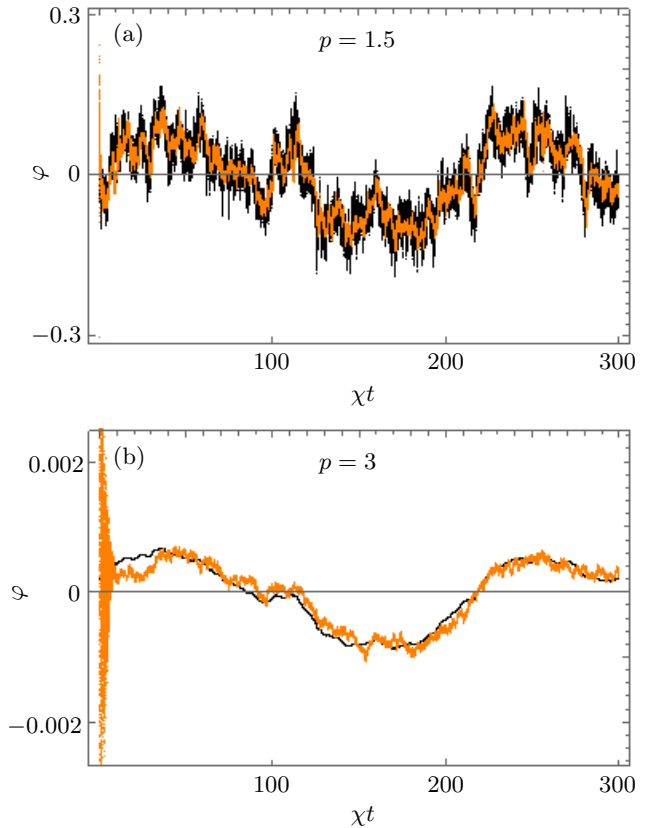

\begin{figure}[tbh]
\centering
\includegraphics[scale=0.7]{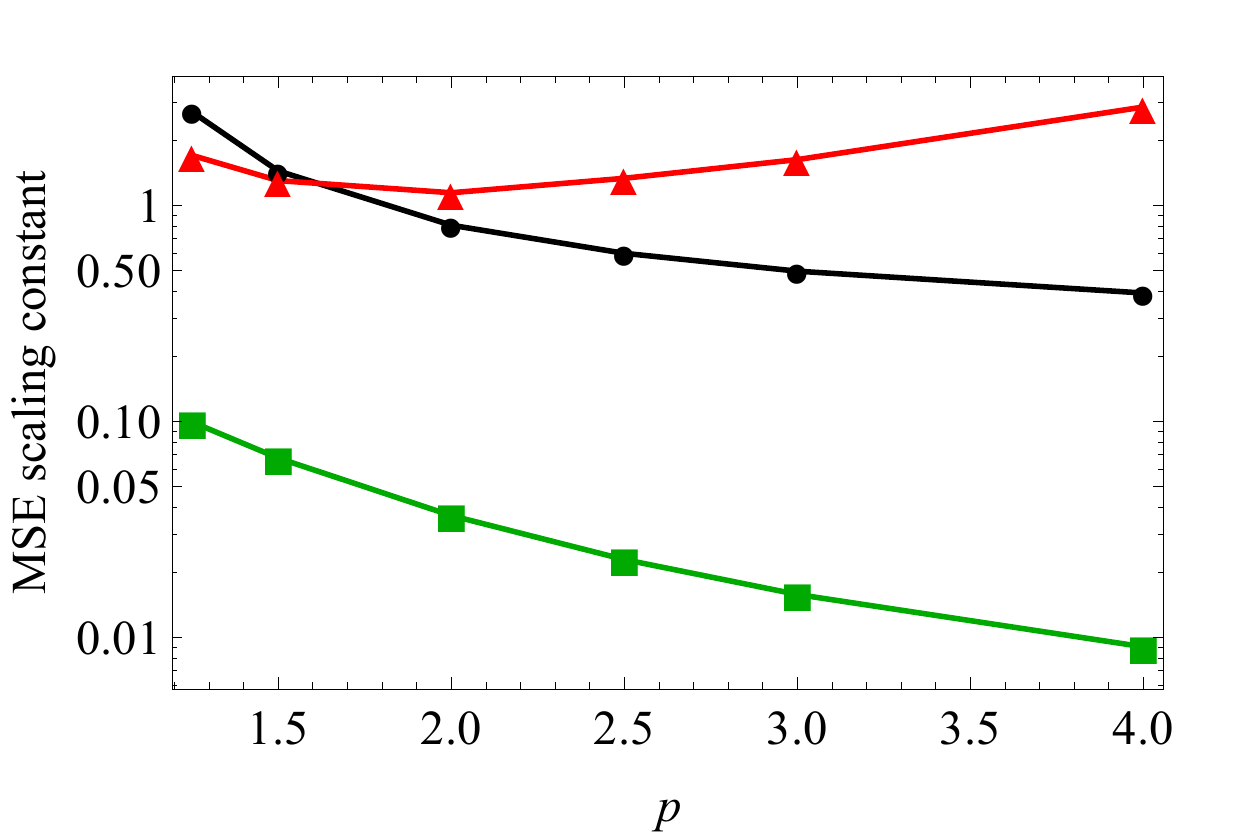}
\caption{The scaling constant of the MSE for the Heisenberg limit (green line with $\blacksquare$), Eq.~\eqref{CHL}, the pulsed measurement (black line with \textbullet), Eq.~\eqref{CA}, and the homodyne scheme with a continuous squeezed state (red line with $\blacktriangle$). }
\label{fig-mfactor}
\end{figure}

\begin{figure}[tbh]
\centering
\includegraphics[scale=0.7]{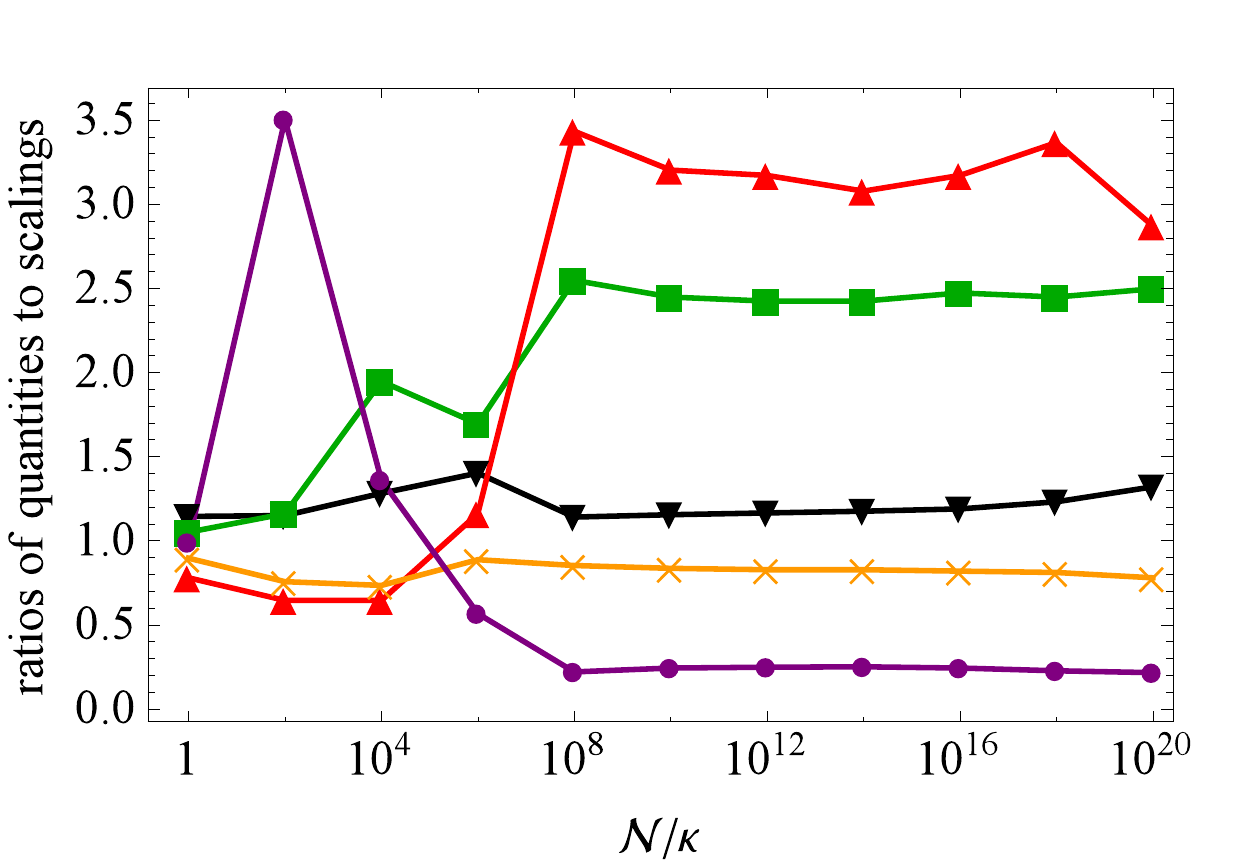}
\caption{The optimal values of various parameters for the Wiener process, i.e.~$p=2$. Black line with $\blacktriangledown$: $\sigma^2/\left({\kappa/\mathcal{N}}\right)^{2/3}$, green line with $\blacksquare$: ${\left(\gamma/\kappa\right)}/{\left({\mathcal{N}/\kappa}\right)}^{5/6}$, red line with $\blacktriangle$: ${\left(\chi/\kappa\right)}/{\left({\mathcal{N}/\kappa}\right)}^{2/3}$, yellow line with $\times$: $e^r/{\left(\mathcal{N}/\kappa\right)}$, and purple line with \textbullet: $\delta/{\left({\kappa/\mathcal{N}}\right)^{1/4}}$. }
\label{fig-p2}
\end{figure}

\section{Numerical results}\label{sec-numerics}
For each value of $\mathcal{{N}}/\kappa$ we have found the minimum MSE by a numerical search for the four parameters $\gamma/\kappa$, $\chi/\kappa$, $\delta$, and $e^r$. In order to do this, we systematically incremented the value of each parameter in turn to find the values that give the minimum MSE.

For $\mathcal{{N}}/\kappa < 5 \times 10^7$ we used a modified form of MSE based on the Holevo variance \cite{Holevo}
\begin{equation}\label{HolevoMSE}
{\rm Re} \left[ \frac{1}{M}\sum_{j = 1}^M e^{i(\breve \varphi _j - \varphi_j)} \right]^{ - 2} - 1\, ,
\end{equation}
where $M$ is the number of samples of the phase estimate.
The Holevo variance is a measure of variance that is naturally modulo $2\pi$, and so is appropriate for phase.
This form of the Holevo variance is analogous to the MSE, because it compares the estimates to the actual values of the phase.
For $\mathcal{N}/\kappa \ge 5 \times 10^7$, we just used the formula for the standard MSE
\begin{equation}
\frac{1}{M}\sum\limits_{j = 1}^M {{\left({\breve\varphi_j-\varphi_j}\right)}^2}\, .
\end{equation}
In this parameter regime, the Holevo MSE is very close to the standard MSE, but performing calculations with Eq.~\eqref{HolevoMSE} is less accurate due to roundoff error.
In the following, we just call both forms the MSE for simplicity.

To calculate the integrals \eqref{eq-A} and \eqref{eq-B}, one can approximate $I$ and $\theta$ as being constant in each interval $[t,t+\Delta t)$, and use the difference equations
\begin{align}
A{\left( {t + \Delta t} \right)} &\approx e^{- \chi \Delta t}A{\left( t \right)} + \frac 1{\chi}\left(1-e^{- \chi \Delta t}\right)I{\left( t \right)}{e^{i\theta }}\, , \\
B{\left( {t + \Delta t} \right)} &\approx e^{- \chi \Delta t} B{\left( t \right)} - \frac 1{\chi}\left(1-e^{- \chi \Delta t}\right){e^{2i\theta }}\, .
\end{align}
In our calculations we made the further approximation that $e^{-\chi \delta t} \approx 1- \chi \delta t$, which simplifies the difference equations to
\begin{align}
A{\left( {t + \Delta t} \right)} &\approx {\left( {1 - \chi \Delta t} \right)}A{\left( t \right)} + I{\left( t \right)}{e^{i\theta }}\Delta t\, , \\
B{\left( {t + \Delta t} \right)} &\approx {\left( {1 - \chi \Delta t} \right)} B{\left( t \right)} - {e^{2i\theta }}\Delta t\, .
\end{align}
We use time steps of $\Delta t=1/{\left(10^3 \chi\right)}$, in which case the approximation $e^{-\chi \Delta t} \approx 1- \chi \Delta t$ is accurate.
Any inaccuracy in the calculation of $A$ and $B$ does not introduce inaccuracy into the simulation as a whole;
instead it means that we are accurately simulating measurements where $A$ and $B$ are calculated in that way.
To give the system of equations time to reach its steady state, we ran the simulations up to time $100/\chi$ without sampling the error.  We then calculated the MSE by sampling the error for every time step up to $300/\chi$. Even though the error was sampled every time step, the samples are strongly correlated for times below $1/\chi$.  Therefore the number of independent samples is effectively the multiple of $1/\chi$ used for the time. We performed $64$ independent integrations from time 0 to $300/\chi$. Therefore, the effective number of independent samples of the error was $12800$. This includes those from different times within one integration.

We found the values of the parameters $e^r$, $\chi$, $\gamma$ and $\delta$ that give the minimum MSE for a range of values of $p$.
For the case of $\delta$, we performed a linear regression of $\log\delta$ versus $\log(\kappa/{\cal N})$ to find the power in the scaling for each value of $p$.
It was found that the powers were consistent with
\begin{equation}
\delta  \sim {\left( {\kappa /\mathcal{N}} \right)^{(p - 1)/(p + 2)}}\, .
\end{equation}
For $p=2$ the scaling corresponds to that found in \cite{DominicE06}.

The ratio of $\delta$ to ${\left( {\kappa /\mathcal{N}} \right)^{(p - 1)/(p + 2)}}$, as well as the ratios of $e^r$, $\chi/\kappa$, and $\gamma/\kappa$ to their predicted scalings in Eqs.~\eqref{er-eq}, \eqref{chi-eq2}, and \eqref{gamsca}, are shown in Fig.~\ref{fig-4scalings}.
The ratio of the minimum MSE to the scaling in Eq.~\eqref{MSE-eq} is shown in Fig.~\ref{fig-var}.
In each case the results are plotted for a range of values of $p$.
For the case of $p=4$, it was not possible to push $\mathcal{ N}/\kappa$ to large values. This is due to the rapid decrease of the MSE for large values of $p$ and the resulting roundoff error in the simulations.

As these results are shown as a ratio to the predicted scalings, if these predicted scalings were exact then the results would all be horizontal lines.
The horizontal lines need not coincide, because the multiplicative constants will be functions of $p$.
It can be seen from these figures that the results are consistent with these scalings, although the scalings are not exact.
The most important results are those shown in Fig.~\ref{fig-var}, which demonstrate that the Heisenberg scaling is obtained for the MSE, with
multiplying factors in the range $1$ to $3$.

There are some discrepancies from straight lines in Fig.~\ref{fig-var}, particularly with the point for ${\cal N}/\kappa=100$ and $p=4$.
The reason for the discrepancy with that point is likely that it takes larger values of ${\cal N}/\kappa$ for the scaling law to be accurate.
There are some smaller discrepancies for ${\cal N}/\kappa=10^6$ for $p=3$ and $p=4$, where the points are noticeably above the neighboring points.
This variation is likely due to chance, because these are Monte Carlo simulations.
These discrepancies are small compared to the overall range of the MSE, which is many orders of magnitude.

In practice it is not possible to use arbitrarily large squeezing;
the current record for squeezing is 15 dB \cite{squeeze}.
Due to the scaling for $e^r$ given in Eq.~\eqref{er-eq}, the optimal amount of squeezing increases with $p$.
For the smallest value of $p$ considered, $p=1.25$, 15 dB is reached for the maximum value of ${\cal N}/\kappa$ shown in Fig.~\ref{fig-4scalings}, so the entire range could be accessed experimentally.
For the other values of $p$, the maximum values of ${\cal N}/\kappa$ would be around $10^8$ ($p=1.5$), $10^5$ ($p=2$), $2\times 10^4$ ($p=2.5$), $8\times 10^3$ ($p=3$), and $4\times 10^3$ ($p=4$).

In Fig.~\ref{fig-cnest} we have plotted the system phase and the estimated phase obtained based on Eq.~\eqref{eq-phiest} for $p=1.5$ and $p=3$. The initial period of transience of the phase estimate can be seen in this figure.
The phase estimate is initially far from the system phase but as we obtain more information from the measurements it locks onto the system phase and follows its variation quite well.

In Fig.~\ref{fig-mfactor} we have compared the scaling constant of the proposed scheme in this paper with the Heisenberg limit, and the pulsed measurement proposed in Ref.~\cite{DominicPRX}. For large values of $\mathcal{N}$ the scaling constant of the Heisenberg limit of Ref.~\cite{DominicPRX} is
\begin{equation}\label{CHL}
c_{Z}=\frac{11}{420}{\left({\frac{p_3}{4}}\right)}^{2/(p+1)}{\left(\frac{1}{4\pi\lambda}\right)}^{2(p-1)/(p+1)},
\end{equation}
with $\lambda\approx0.7246$ and $p_3=(p+1)(p+2)(p+3)$. The scaling constant for the MSE achievable by the pulsed  method of \cite{DominicPRX} is 
\begin{align}\label{CA}
c_{A}=\frac{p+1}{p-1}{\left({4 |z_A|^3/27}\right)}^{(p-1)/(p+1)}\pi^{2p/(p+1)}\, ,
\end{align}
with $z_A\approx -2.338$.
As can be seen from this figure, the pulsed method (with assumed ideal phase measurements) performs better than the continuous squeezing method (with adaptive homodyne measurements) for larger $p$. 

We have also compared our results using the improved numerical techniques to those in Ref.~\cite{DominicE06} for the case of the Wiener process, i.e.~$p=2$.
Our new results are slightly different, although they are qualitatively similar in that they follow the predicted scalings.
The results are plotted together in a single graph in Fig.~\ref{fig-p2}.
This figure shows the same quantities as Fig.~3 in Ref.~\cite{DominicE06}.

\section{Conclusion}
In this work we investigated estimation of a time-varying phase in an adaptive homodyne scheme using a continuous squeezed state.
We considered a phase with time-invariant Gaussian statistics and power-law spectral density.
We showed that assuming it is possible to achieve arbitrarily high squeezing, this scheme gives Heisenberg scaling for the MSE in the phase estimate.
Moreover, we found that for $p\le 1.5$ the scaling constant obtained with the adaptive method is smaller than the scaling obtained with the sampling method proposed in Ref.~\cite{DominicPRX}.
For larger values of $p$ the scaling constant is larger than that for the method of \cite{DominicPRX}.
We also recalculated the optimal values of the parameters for the Wiener process and gave more accurate results for this case.

Although we have obtained Heisenberg scaling for the MSE, there is still the possibility of improvements in the scaling constant.
An obvious way to obtain an improvement in the scaling constant is to use smoothing, where data from before and after a particular time is used to estimate the phase at that time.
It can be expected that the reduction in the MSE from smoothing is about a factor of 2, provided there are not significant correlations between the errors before and after the time of interest.
There is also the potential for obtaining better results using a different analysis of the data better taking into account its correlations, for example Kalman filtering \cite{Mankei}.
It is also possible that an approach using the Bayesian probability distribution might give improved results.

\acknowledgments
We acknowledge helpful discussions with Howard Wiseman.
DWB is funded by an Australian Research Council Future Fellowship (FT100100761) and an Australian Research Council Discovery Project (DP160102426).

\end{document}